\documentclass[useAMS,usenatbib]{mn2e}

\usepackage{graphicx,color}

\title[Red giant pulsation modes, masses and evolution]{The pulsation modes, masses and evolution of luminous red giants}
\author[P. R. Wood]{P. R. Wood\thanks{E-mail: peter.wood@anu.edu.au}\\
Research School of Astronomy and Astrophysics, Australian National University, Cotter Road, Weston Creek ACT 2611, Australia}

\begin{document}

\date{}

\pagerange{\pageref{firstpage}--\pageref{lastpage}} \pubyear{2014}

\maketitle

\label{firstpage}

\begin{abstract}

The period--luminosity sequences and the multiple periods of luminous
red giant stars are examined using the OGLE III catalogue of
long-period variables in the Large Magellanic Cloud.  It is shown that
the period ratios in individual multimode stars are systematically
different from the ratios of the periods at a given luminosity of
different period--luminosity sequences.  This leads to the conclusion
that the masses of stars at the same luminosity on the different
period--luminosity sequences are different.  An evolutionary scenario
is used to show that the masses of stars on adjacent sequences differ
by $\sim$16--26\% at a given luminosity, with the shorter period
sequence being more massive.  The mass is also shown to vary across
each sequence by a similar percentage, with the mass increasing to
shorter periods.  On one sequence, sequence B, the mass distribution
is shown to be bimodal.  It is shown that the small amplitude
variables on sequences A$^\prime$, A and B pulsate in radial and
nonradial modes of angular degree $\ell$=0, 1 and 2, with the $\ell$=1
mode being the most common.  The stars on sequences C$^\prime$
and C are predominantly radial pulsators ($\ell$=0).  
Matching period ratios to pulsation models shows
that the radial pulsation modes associated with sequences A$^\prime$,
A, B, C$^\prime$ and C are the 4th, 3rd, 2nd and 1st overtones and the
fundamental mode, respectively.

\end{abstract}


\section{Introduction}

It is now well known that luminous red giants are variable and that
the periods of the observed variability lie on discrete sequences in
period--luminosity diagrams.  \citet{woo99} and \citet{woo00} used
data from MACHO observations of stars in the Large Magellanic Cloud
(LMC) to define five sequences A, B, C, D and E in the ($I_{\rm W}$,
$\log P$) and ($K$, $\log P$) planes, respectively (where $I_{\rm W}$
is a reddening-free Weisenheit index defined as $I$--1.38($V$--$I$)).
The sequences A, B and C were identified as containing pulsating
stars, while sequence E was found to consist of binary stars and
sequence D was the sequence formed by the long secondary periods
(LSPs) that occur in 30--50\% of luminous pulsating red giants.  In
this paper, the pulsating stars will be discussed.  These stars lie
on sequences A, B and C and two other sequences A$^\prime$
and C$^\prime$ which are described below.  Further examination of the ellipsoidal
variables on sequence E can be found in \citet{sos04b},
\citet{nic10}, \citet{nic12} and \citet{nie14} while studies of the
origins of the LSPs can be found in \citet{hin02}, \citet{oli03},
\citet{wok04}, \citet{nic09}, \citet{sto10} and \citet{tak15}.

Following the discovery of sequences A, B and C, \citet{ita04} used
extensive $K$ band photometry combined with MACHO periods to show that sequence
B was in fact two sequences which were renamed B and C$^\prime$ .  The
next refinement to the luminous sequences was by \citet{sos04a} who
used OGLE data to show the existence of a new sequence on the short
period side of sequence A.  \citet{sos04a} labelled this sequence
a$_{\rm 4}$ but in this paper the designation A$^\prime$ will be used
\citep[e.g.][]{der06,tab10,sos13b}.  \citet{sos04a} also made the important
discovery that sequence A consisted of three closely-spaced parallel
sequences.  \citet{sos07} claimed that sequences B and A$^\prime$ also
consisted of three closely-spaced parallel sequences (hereinafter referred
to as sub-sequences).  More recently,
\citet{sos13a}  have drawn attention to a probable faint sequence
designated F between sequences C$^\prime$ and C.

\citet{woo99} assumed that all the variable luminous red giants found
by the MACHO observations were asymptotic giant branch (AGB) stars.
However, \cite{ita02} and \citet{kis03} showed that both AGB and red
giant branch (RGB) stars were variable and that there was a slight
offset in period between the sequences of RGB and AGB stars.
\citet{sos04a} also noted this offset but they did not find evidence
for two slightly offset sequences for sequence A$^\prime$ which they
suggested consisted of AGB stars only.  However, this result was
challenged by \citet{tab10} who showed that an RGB component of
sequence A$^\prime$ does exist.

Various attempts have been made to identify the pulsation modes
associated with the pulsation sequences C, C$^\prime$, B, A and
A$^\prime$ and the sub-sequences of B, A and A$^\prime$.
\citet{woo99} (see also \citealt{woo96}) showed that the radial
fundamental mode could fit sequence C with reasonable assumptions
about the mass of the star and that the first few radial overtone
modes had periods similar to those of sequences B (which now includes
C$^\prime$) and A, although the data available at the time was not
extensive enough for a detailed matching of these latter sequences
with modes.  They also noted that the large-amplitude
Mira variables fell on sequence C, making
them fundamental mode pulsators.  Additional evidence in favour
of this modal assignment is provided by the fact that nonlinear fundamental mode
radial pulsation models reproduce the light and velocity
curves of local Mira variables quite well \citep[e.g.][]{bes96,ire08}. 
Recently, using extensive
observational data from OGLE, \citet{sos13a} showed that the period
ratios in stars of sequences C and C$^\prime$ were consistent with
those expected for fundamental mode pulsation of sequence C and first
overtone pulsation on sequence C$^\prime$.  \citet{tak13} compared
period ratios for OGLE variables near the tip of the RGB with models
and concluded that sequences C$^\prime$, B and A corresponded to the
radial first, second and third overtones, consistent with the
assignments by \citet{woo99} and \citet{sos13a}.

In contrast to the above results, \citet{sos07}
concluded that the sequences A, B and C$^\prime$
corresponded to the radial 2nd overtone, first overtone and
fundamental modes, respectively, when they used used radial
pulsation models to try fit the sequences in the period--luminosity
(PL) plane at RGB luminosities (see also \citealt{dzi10}).  This
means that the longer period sequence C containing the Miras has no
explanation in terms of radial pulsation.  This study was extended by
\citet{mos13} who matched the A and B sequences at RGB luminosities to
sequences of Kepler radial mode pulsators in a PL plane where
frequency data was used as a proxy for luminosity.  This result
produces a close match between sequences B and A and the radial first
and second overtones, respectively, in agreement with \citet{sos07}
but not with \citet{woo99}, \citet{sos13a} and \citet{tak13}.  There
thus exists two extant assignments of the radial modes to the PL
sequences A, B, C$^\prime$ and C that differ by one radial order.

In another study using Kepler data, \citet{ste14} showed that in the
luminous Kepler red giants, triplets of modes with periods
corresponding to spherical degrees $\ell$=0, 2 and 1 (in increasing order
of period) are excited for a given radial order.  The mode with
$\ell$=1 has the largest amplitude for radial orders $n$=3 or more
(the second or higher radial overtones).  This suggests the OGLE
sequences B, A and A$^\prime$ may be dominated by $\ell$=1 modes.

In this paper, the period and luminosity data given in the OGLE III
catalogue of long-period variables (LPVs) in the LMC \citep{sos09} is
analysed in detail.  Purely empirical results related to the modes,
masses and evolution of the stars on the sequences are derived.  Then
the data are compared to pulsation models in order to throw
further light on the modes of pulsation involved.

\section{The sequences of luminous red giants}\label{seqs}

\begin{figure}
\includegraphics[width=1.0\columnwidth]{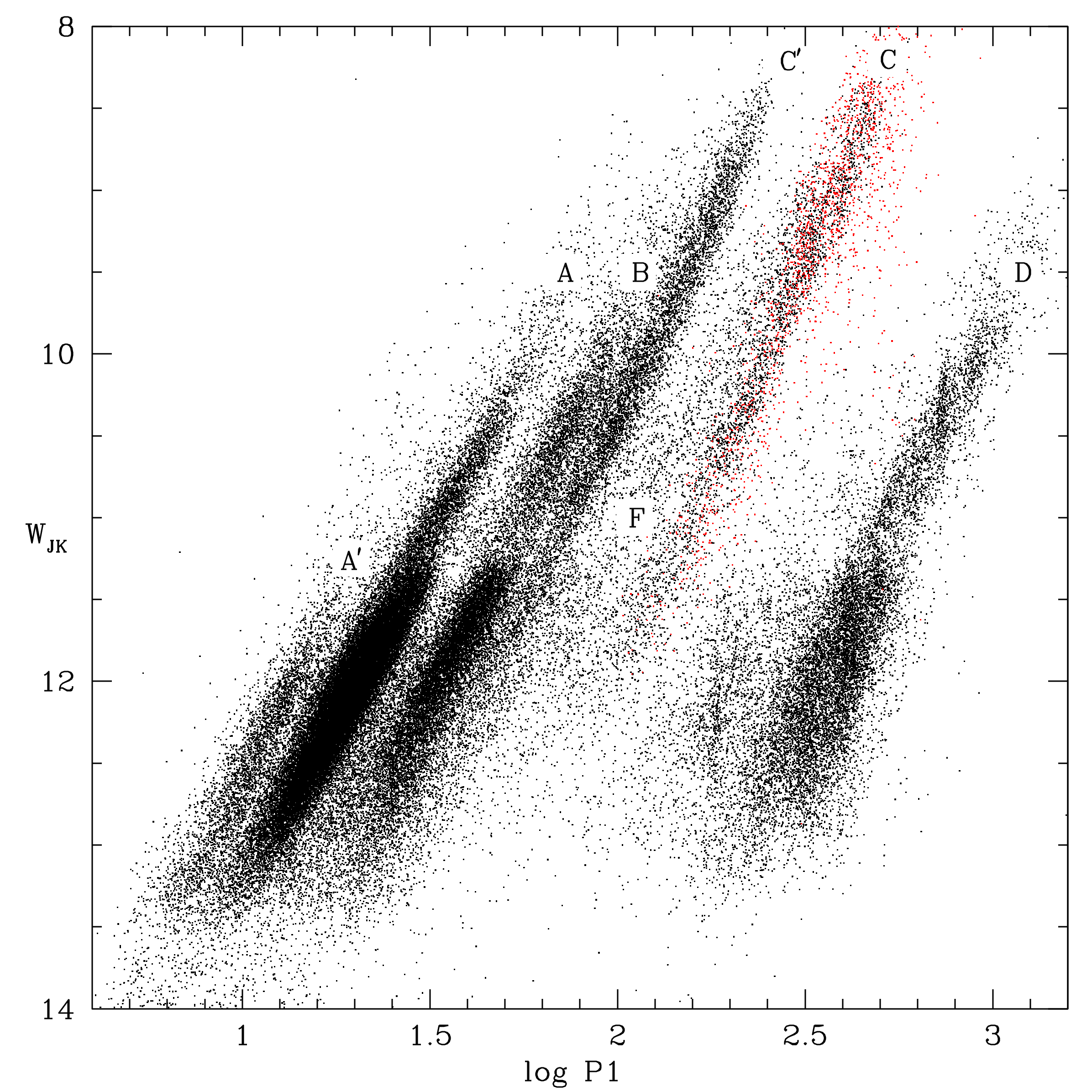}
\caption{The sequences of variable luminous red giants in the ($W_{\rm JK}$,$\log P1$)
plane, where $P1$ is the period of the mode with the largest amplitude and 
$W_{\rm JK} = K - 0.686(J-K)$ is a reddening-free measure of the luminosity.  
Mira variables are shown as red points (in the online version of the paper).  
The ellipsoidal variables belonging to sequence E are not shown.}
\label{wjk-p}
\end{figure}

The sequences of luminous red giants in the LMC are shown in
Figure~\ref{wjk-p}.  The period data for this figure come from the
OGLE III catalogue of LPVs in the LMC \citep{sos09}.  All the
variables in the catalogue are included in the analysis in this paper
- no selection is made according to the catalogue types Mira,
SRV or OSARG.  In the catalogue, three periods are given for each star
with the first (or primary) period (P1) having the largest amplitude
and the secondary periods P2 and P3 having progressively smaller
amplitudes (most luminous red giants exhibit multiple periods of
variation).  The luminosity in Figure~\ref{wjk-p} is represented by
$W_{\rm JK} = K - 0.686(J-K)$ where the $J$ and $K$ magnitudes are
from the 2MASS point source catalogue \citep{skr06}.  The individual
sequences are labelled in Figure~\ref{wjk-p} although sequence E
(produced by the binary ellipsoidal variables) is omitted as these
stars are not included in the catalogue. The tip of the RGB is clearly
seen at $W_{\rm JK} \sim 11.3$.

The large-amplitude Mira variables are shown as red points in
Figure~\ref{wjk-p}.  They clearly belong to sequence C.  As noted in
the Introduction, there are several lines of evidence which suggest
that stars on sequence C are fundamental mode radial pulsators, although some
authors have provided arguments against this modal assignment.  In
Section~\ref{comp_mods}, we will present what we believe to be 
compelling evidence that the stars on sequence C are indeed fundamental mode
pulsators.  For these reasons, and to simplify some of the discussion,
in this paper it will be assumed from here on that the sequence C
stars, including the Mira variables, are fundamental mode radial pulsators.

It is worth considering at this point what causes the existence of the
sequences in diagrams like Figure~\ref{wjk-p}.  Firstly, remember that
the period used in Figure~\ref{wjk-p} is the primary or
largest-amplitude period.  At a given luminosity, a star will have
possible modes of pulsation with periods that could fall on any of the
sequences or (as shall be shown below) between the sequences.  The
actual position of a star in the figure is determined by the period of
the mode of largest amplitude.  For self-excited modes, this mode
is the one that has the strongest driving - see \citet{sos13a} for a
demonstration of this effect for 1st overtone pulsation on sequence C$^\prime$.
For modes that are excited stochastically by convective motions
(solar-like oscillations), there is also a preferred pulsation period
which is related to the acoustic cut-off frequency
\citep[e.g.][]{dzi01,mos13,ste14}.

\section{Period ratios}

\begin{figure*}
\includegraphics[width=0.8\textwidth]{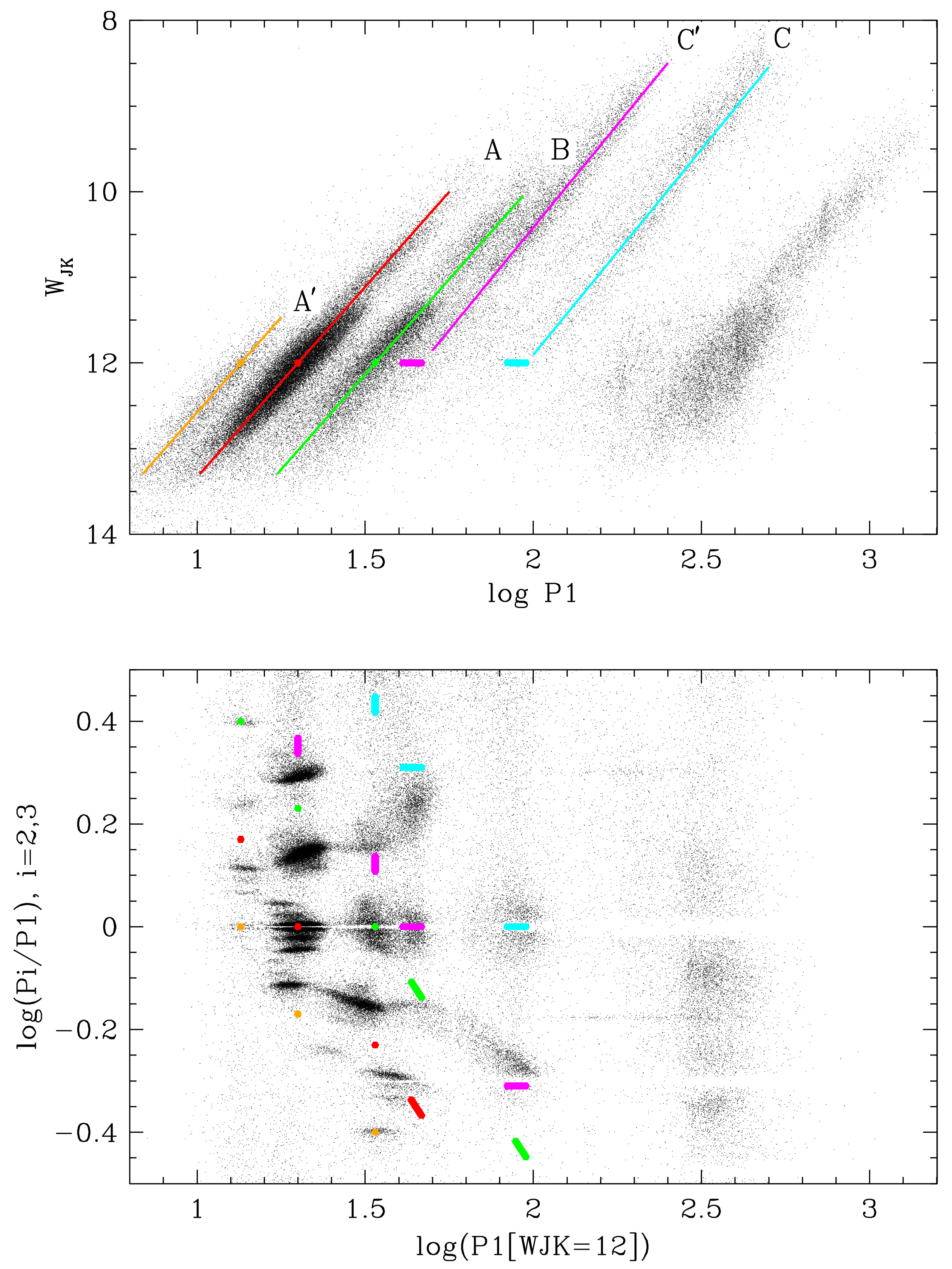}
\caption{
Top panel: The black points show the stars in the ($W_{\rm 
JK}$, $\log P1$) plane while the coloured lines are fits to the sequences.
The coloured dots and thick coloured lines are projections of the coloured lines 
onto $W_{\rm JK} = 12$ along lines of slope $dW_{\rm JK}/d\log P1 = 4.444$.
Bottom panel: The black points show the logarithmic period ratios $\log P2/\log P1$
and $\log P3/\log P1$ in individual stars plotted
against $\log P1$[$W_{\rm JK}$=12] (see text for the definition of this quantity).  
The coloured points and short lines correspond to the ratios
of the periods on the fit lines shown in the upper panel (see text for
full details).  The colours can be seen in the online version of the paper.
}
\label{em_strips}
\end{figure*}

Period--luminosity (PL) laws such as those in Figure~\ref{wjk-p}
provide one type of diagram for the comparison of models and observations of
variable stars.  A more precise diagram for comparing observations and 
models is the Petersen diagram wherein period ratios in multimode pulsators
are plotted against the longer of the two periods in the ratio.  In this section,
a modified Petersen diagram appropriate for LPVs is discussed.

The upper panel of Figure~\ref{em_strips} shows the
same PL diagram as Figure~\ref{wjk-p} but now coloured straight lines
have been fitted to each of the pulsation sequences C, C$^\prime$, B,
A and A$^\prime$.  It has been assumed that the three sequences B, A and
A$^\prime$ have the same slope $dW_{\rm JK}/d\log P1 = 4.444$ while
the two sequences C and C$^\prime$ have a slightly different slope 
$dW_{\rm JK}/d\log P1 = 4.800$.

Since many of the variable LPVs are multimode pulsators, 
they can be plotted in a Petersen diagram.  
The lower panel of Figure~\ref{em_strips} shows a variant of this
diagram where the logarithmic period ratios $\log P2$/$P1$ and 
$\log P3$/$P1$ in each star are plotted against the
variable $\log P1$[$W_{\rm JK}$=12].  The latter quantity is the
value of $\log P1$ obtained by projecting the actual value of $\log
P1$ along a line parallel to sequence A until it meets the level $W_{\rm JK} = 12$.
Specifically, $P1$[$W_{\rm JK}$=12] = $\log P1 + (W_{\rm JK} - 12)/4.444$.
Stars with primary periods on a given sequence are thus bounded in the
same horizontal range in the lower panel of Figure~\ref{em_strips}, with this range corresponding
to the values of $\log P1$ for this sequence when $W_{\rm JK}=12$.  
For LPVs, this form of Petersen diagram is much better at revealing preferred
period ratios than the traditional Petersen diagram used by \citet{sos04a}
and \citet{tak13} which plots
$P_{\rm short}/P_{\rm long}$ against $P_{\rm long}$.  A form of Petersen diagram
which used a quantity like $P1$[$W_{\rm JK}$=12] on the horizontal axis
was presented by \citet{sos07}.

The coloured points and short lines in the lower panel of
Figure~\ref{em_strips} correspond to the ratios of the periods on the
fit lines shown in the top panel i.e. they represent the ratios of the
periods of the different period--luminosity sequences.  The horizontal
position of the coloured point/lines is $\log P1$[$W_{\rm JK}$=12] for
the line which is in the denominator of the period ratio.  The colour
of the point/line is the colour of the line in the numerator.
Coloured points correspond to period ratios of the fits to sequences
A$^\prime$, A and B (these all have the same slope).  For example, the
red point at ($\log P1$[$W_{\rm JK}$=12], $\log Pi$/$P1$) $\approx$
(1.12, 0.17) represents the ratio of $\log P1$ at the middle of
sequence A to $\log P1$ at the middle of sequence A$^\prime$, the red
point near (1.3, 0.0) represents this ratio for sequence A to itself
and the red point near (1.52, -0.23) represents the ratio for sequence
A to sequence B.  Since the lines fitted to sequences C and C' have a
slightly different slope to that of sequence A, the period ratios of
fits involving sequences C or C' lead to the short coloured lines.
For example, the short red line at ($\log P1$[$W_{\rm JK}$=12], $\log
Pi$/$P1$) $\approx$ (1.65, -0.35) represents the ratio of $\log P1$ at
the middle of sequence A to $\log P1$ at the middle of sequence
C$^\prime$.

We now examine the structure in the lower panel of Figure~\ref{em_strips} in detail.

\subsection{Observed period ratios for individual stars and for sequences}\label{pr}

The black points in the lower panel of Figure~\ref{em_strips} show observed period ratios 
for individual stars.  The clustering of these points represents ratios of specific
modes present in the data.  Firstly, we start with the points in the vertical strip
1.2 $\la$ $\log P1$[$W_{\rm JK}$=12] $\la$ 1.4 which belong to stars with their
primary period on sequence A.  There is a large cluster of points near
($\log P1$[$W_{\rm JK}$=12], $\log Pi$/$P1$) = (1.3, 0.14) which one would expect to belong
to stars on sequence A that show a secondary period on sequence B.  Similarly,
the cluster of points near
($\log P1$[$W_{\rm JK}$=12], $\log Pi$/$P1$) = (1.3, 0.3) should correspond
to stars on sequence A that show a secondary period on sequence C$^\prime$
and a third cluster of points near
($\log P1$[$W_{\rm JK}$=12], $\log Pi$/$P1$) = (1.3, -0.12) which should correspond
to stars on sequence A that show a secondary period on sequence A$^\prime$.
However, the period ratio of sequences B, C$^\prime$ and A$^\prime$ to sequence A,
represented by the green, magenta and orange points at $\log P1$[$W_{\rm JK}$=12]=1.3,
respectively, do \emph{not} coincide with these clouds of black points.  It can be
seen that this problem exists for stars with primary periods on any of the sequences.
We will explore this situation in detail in Section~\ref{chans}.  

Another interesting feature of Figure~\ref{em_strips} is that there are clouds
of points such as those near ($\log P1$[$W_{\rm JK}$=12], $\log Pi$/$P1$) = (1.45, -0.15)
for which the primary period lies \emph{between} the sequences.  One can see in the upper
panel of Figure~\ref{em_strips} that some stars do have their primary periods
between the sequences although they are not as common as the stars on the sequences.
In fact, it turns out that the existence of stars between the sequences is related to the problem noted
in the previous paragraph and we come back to this in Section~\ref{chans}.

\subsection{Closely spaced periods within a sequence}\label{pr1}

A prominent feature in the lower panel of Figure~\ref{em_strips} is the
set of sub-ridges first noted by \citet{sos04a} and especially prominent for
stars on sequence A where five almost-horizontal sub-ridges can be seen in the cluster of points near 
($\log P1$[$W_{\rm JK}$=12], $\log Pi$/$P1$) = (1.3, 0.0).  These five sub-ridges
must be due to pulsation modes with very closely spaced periods.  The spacing is 
sufficiently small that the individual modes can not be distinguished in
a PL diagram such as the top panel of Figure~\ref{em_strips}.  The period separation
of low-order radial modes in red giants is much larger than the period separation
of the modes belonging to the five sub-ridges of sequence A so nonradial pulsation modes,
as well as a radial pulsation mode, must be involved. 

The five sub-ridges of sequence A can be created
by three closely-spaced modes each separated in $\log P$ by $\sim$0.022, provided that
any of the three modes can be the primary mode in at least some stars.  We assume here
that this is the case (if one mode was always dominant, five modes would be required).
There is some evidence that there are three sub-ridges, corresponding to two close modes,
associated with sequence B (see the cluster of points around 
($\log P1$[$W_{\rm JK}$=12], $\log Pi$/$P1$) = (1.53, 0.0)).  This evidence will be
made more apparent in Section~\ref{ldep_pr}.

As noted above, the three modes associated 
with sequence A must include nonradial modes because
of their close spacing.  We assume that all stars on sequence A, and
indeed all stars on any given sequence, have the same radial order of
pulsation.  Nonradial pulsation models of luminous red giants
\citep[e.g.][]{dzi01,mos13,ste14} show that for modes of a given
radial order but different angular degree $\ell=0$, 1 or 2 (the only
ones likely to be observed in light curves), the $\ell=1$ mode has the
longest period, the $\ell=0$ mode has the shortest period and the
$\ell=2$ mode has an intermediate period.  

Figure~\ref{em_strips} provides a method for determining which of the
three modes within sequence A is dominant.  Among the stars with their primary
period on sequence A$^\prime$
(the points with 1.05 $\la$ $\log P1$[$W_{\rm JK}$=12] $\la$ 1.2), those that also have the
closely-spaced modes associated with sequence A can be seen near ($\log P1$[$W_{\rm JK}$=12], 
$\log Pi$/$P1$) = (1.12, 0.12).  There are three horizontal bands here with the
same spacing as the sub-ridges associated with sequence A.  The
upper band is by far the most prominent.  Assuming that one mode dominates sequence A$^\prime$
(because there are no prominent sub-ridges near ($\log P1$[$W_{\rm JK}$=12],
$\log Pi$/$P1$) = (1.12, 0.0)),
this band structure suggests that it is the longest period mode on sequence A that
generally has the largest amplitude.  As noted above, this is the $\ell=1$ mode.
The dominance of the $\ell=1$ mode found on sequence A in the present high luminosity
red giants is consistent with the findings of
\citet{ste14} from Kepler observations but different from that of
\citet{mos13} who find the radial $\ell=0$ mode becomes dominant at
high luminosities in their analysis of Kepler stars.

\begin{figure}
\includegraphics[width=1.0\columnwidth]{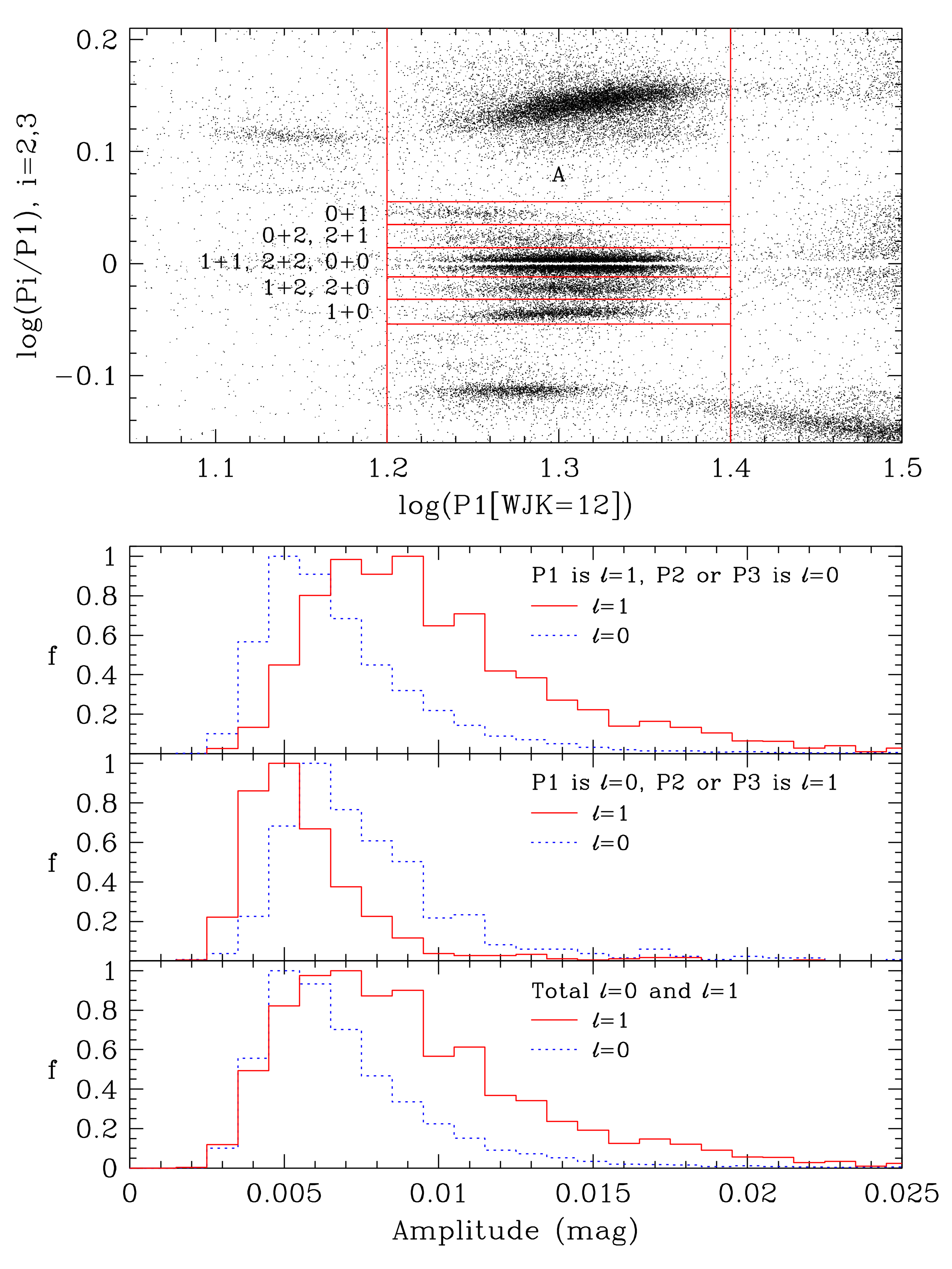}
\caption{Top panel: The small section of the Petersen diagram containing the
five sub-ridges formed by the three pulsation modes forming sequence A.  The five boxes
enclose the five sub-ridges and their names are shown to the left of the boxes.
Bottom panel: The amplitude distribution of the $\ell=0$ and $\ell=1$ modes
of sequence A when these two modes are present together.  See text for 
further details.}
\label{mode_amps}
\end{figure}

\begin{table}
\caption{Stars on the sub-sequences of sequence A}
\centering
\begin{tabular}{l c c}
\hline
Subgroup & No. of stars & Fraction of sequence A\\  
\hline
n(A) & 38526 & 1.000\\ 
n(0+1) & 639 &  0.017 \\ 
n(0+2,2+1) & 1152 &  0.030 \\  
n(1+1,2+2,0+0) & 10866 &  0.282 \\ 
n(1+2,2+0) & 2930 &  0.076 \\ 
n(1+0) & 2754 &  0.071 \\ 
\hline
\end{tabular}
\label{n_subseqA}
\end{table}

Estimates of the relative amplitudes of the three modes that make up the three
sub-sequences of sequence A can be found by examining the populations
of the five sub-ridges of sequence A in the Petersen diagram.  An
exploded view of these sub-ridges is shown in the top panel of
Figure~\ref{mode_amps}.  Boxes have been drawn to enclose the
sub-ridges and the sub-ridges have been named after the possible modes
occurring in them.  Under our assumption that the three modes involved
are the $\ell=0$, 1 and 2 modes of a single radial order and that the
period increases from $\ell=0$ to 2 to 1, a star found in the top box
must have $\ell=0$ for its primary period P1 and $\ell=1$ for the
secondary period (P2 or P3).  This box has been labelled {\tt 0+1}. 
Similarly, the lowest box {\tt 1+0} must have $\ell=1$ for its
primary period P1 and $\ell=0$ for its secondary period.  The central
box is made up of stars where the same mode is detected twice with
slightly different periods and the mode involved could be any of the
$\ell=0$, 1 and 2 modes.  The two intermediate boxes are labelled by
the possible pairs of modes that could be present.  Note that the
period separation $\Delta\log P \approx 0.043$ 
of the $\ell=0$ and $\ell=1$
modes obtained from the positions of the sub-ridges agrees with the period
separation of these modes given by the observations and the models of
\citet{ste14}, giving added support for the mode identifications.

The fact that the ridges in boxes {\tt 0+1} and {\tt 1+0} are not
strictly parallel to the horizontal axis means that $\Delta\log P$
varies systematically by a small amount across sequence A.
Quantitatively, for stars on sequence A, $\Delta\log P$ decreases by
$\sim$0.0045 as $\log P1$[$W_{\rm JK}$=12] increases by 0.1.  An
explanation for this behaviour is given in Section~\ref{prat-m}.

The number of stars in each box is given in Table~\ref{n_subseqA}.
The total number $n({\rm A})$ of stars with their primary period P1 on
sequence A is also given.  This latter number is the total number of
stars with $\log P1$[$W_{\rm JK}$=12] lying between the values
corresponding to the two vertical lines in Figure~\ref{mode_amps}.
Stars that have both the $\ell=1$ and $\ell=0$ modes detected in their
light curves lie in boxes {\tt 1+0} and {\tt 0+1}.  It is clear that
$n({\rm 0+1})$ is much smaller than $n({\rm 1+0})$ which shows that
the $\ell=1$ mode is much more likely to be detected as the
\emph{primary} pulsation mode than the $\ell=0$ mode and this in turn
suggests that the $\ell=1$ mode is likely to have a larger amplitude
than the $\ell=0$ mode.  This is shown directly in the bottom panel of
Figure~\ref{mode_amps} where histograms of modal amplitudes are shown
(the amplitudes come from the OGLE III catalogue).  The top sub-panel
shows the amplitudes of the $\ell=0$ and $\ell=1$ modes for stars in
box {\tt 1+0}, the middle sub-panel shows the amplitudes for stars in
box {\tt 0+1} and the bottom sub-panel shows the sum of the top and
middle sub-panels.  It is clear that the $\ell=1$ mode tends to have
the larger amplitude in stars where the $\ell=0$ and $\ell=1$ modes
are both present although the non-zero population in box {\tt 0+1}
shows that the $\ell=0$ mode can sometimes have a larger amplitude
than the $\ell=1$ mode.  

The amplitude distributions shown in Figure~\ref{mode_amps} apply to
stars with two modes on sequence A.  However, about 80\% of sequence A
stars have only one detected mode on sequence A (see the fractions in
the last column of Table~\ref{n_subseqA}, and note that stars on the
middle sub-ridge have two detections of the same mode).  Assuming that
the relative amplitude distributions shown in Figure~\ref{mode_amps}
are typical, the stars with only a single mode on sequence A will be
mostly $\ell=1$ pulsators, in agreement with the result derived above
from the stars that have their primary mode on sequence A$^\prime$ and
a secondary mode on sequence A.

One factor that has not been mentioned above is the orientation of the
pole of the pulsating star relative to the observer in the case of
nonradial pulsation.  Different orientations will cause different
apparent amplitudes of pulsation for a nonradial pulsator, with a zero
apparent amplitude being possible for specific values of $\ell$, $m$ and the
orientation.  As a result, the true numbers of $\ell=1$ and $\ell=2$
nonradial pulsators relative to the number of $\ell=0$ radial
pulsators should be higher than indicated by the numbers in
Table~\ref{n_subseqA}.

\section{Multiple modes in the PL diagram}\label{chans}

\begin{figure*}
\includegraphics[width=0.93\textwidth]{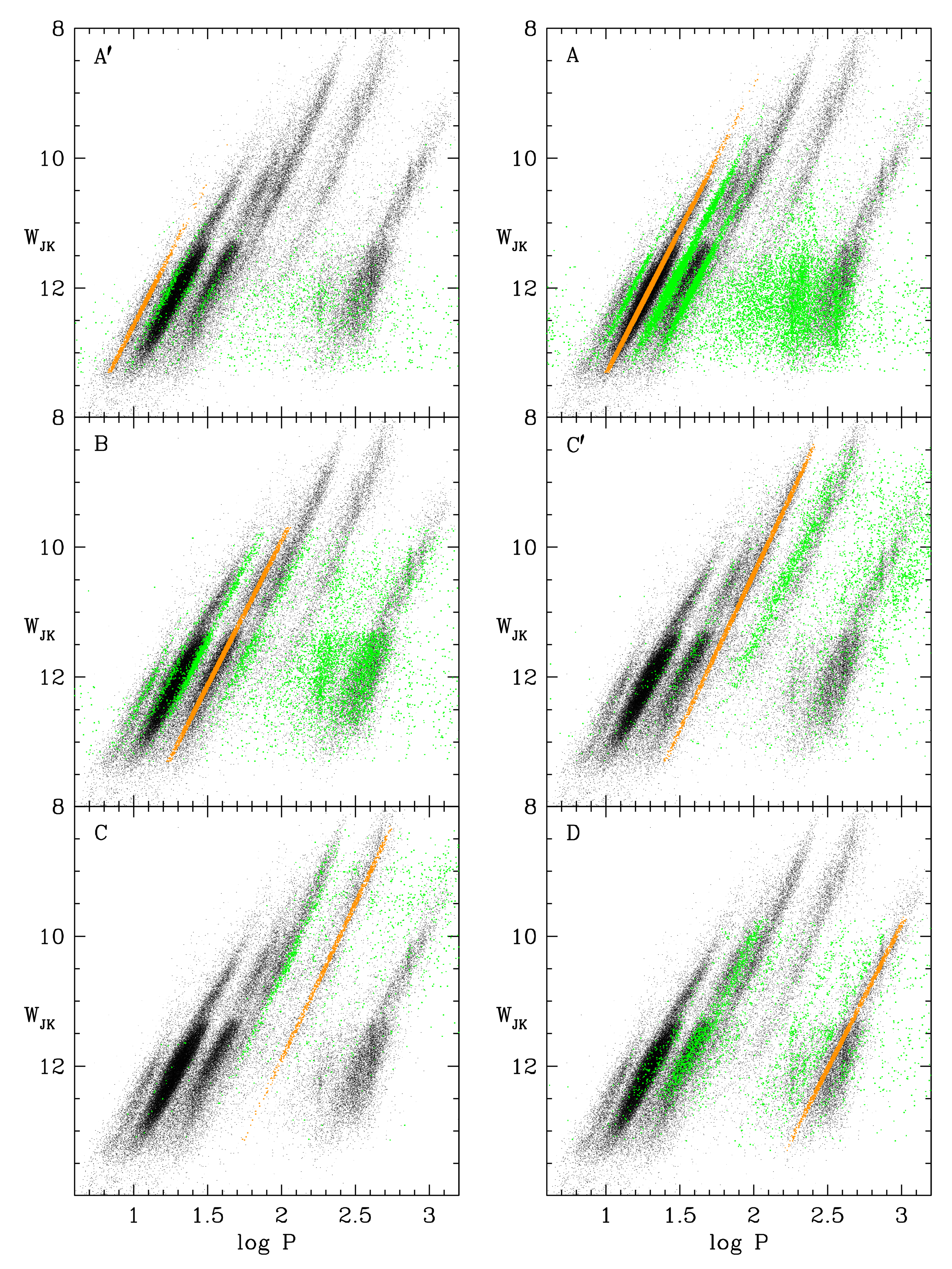}
\caption{
Plots of $W_{\rm JK}$ against the primary period $P1$ for stars in the
OGLE III catalogue (black points).  Each panel is labelled by a
sequence name and the stars with primary periods lying in a channel
running along the centre of the named sequence have been plotted as
orange points.  Secondary periods $P2$ and $P3$ for the stars with
$P1$ in the orange channel are plotted as green points.  For clarity,
green points close to the orange channel are not plotted: the excluded points have
$\mid \log P2 -\log P_{\rm c}\mid$ $<$ 0.07 and $\mid \log P3 -\log
P_{\rm c}\mid$ $<$ 0.07, where $\log P_{\rm c}$ is $\log P$ at the
centre of the channel.  The excluded points are second detections
of the mode represented by the orange points. 
The colours can be seen in the online version of the paper.
}
\label{channel}
\end{figure*}

In Section~\ref{pr} it was shown that period ratios in individual
stars do not agree with the period ratios of the mean sequences.  A
vivid demonstration of this situation is shown in
Figure~\ref{channel}.  Looking at panel A, it can be seen that the
secondary periods form three strips parallel to sequence A, but they are
closer to sequence A than are sequences A$^\prime$, B or C$^\prime$.
This situation, wherein the secondary modes are closer to the primary
mode than are the adjacent sequences to the central sequence, exists
for sequences A$^\prime$, A and C.  It also exists for the modes
shortward of the primary mode of sequence B and for the mode longward
of the primary mode for sequence C$^\prime$.  The only exception to
this rule occurs for the mode longward of the primary mode of sequence
B (which is further from sequence B than is sequence C$^\prime$) and
for the mode shortward of the primary mode of sequence C$^\prime$
(which is further from sequence C$^\prime$ than is sequence B).
We come back to this exception later.

We now seek to explain why the green strips in Figure~\ref{channel}
do not align with the sequences.  Firstly, it is assumed here that the
pulsation modes that produce the green strips are the same pulsation
modes as those that produce the corresponding,
neighbouring sequences (where the correspondence is obtained from
the ordering in $\log P$ of the strips and sequences).  This
assumption will be justified in Section~\ref{evsec}.
In this situation, the period of the same mode must
be different on a green strip and on its corresponding sequence.
Now, in general, the period of a pulsation mode can be written as $P
\propto R^{\alpha}/M^{\beta}$ where $\alpha$ and
$\beta$ are both positive and for red giants typically range from
1.5--2.3 and 0.5--1.9, respectively \citep[e.g.][]{fw82}.  In
addition, the position of the giant branch at a given luminosity means
that $T_{\rm eff} = T_{\rm eff}(L,M)$.  Combining these two relations
with the definition $L = 4\pi \sigma R^2T_{\rm eff}^4$ yields $P =
P(L,M)$.  Thus if a green strip and
its corresponding sequence have the same mode but a
different $P$ at a given $L$ then the masses of the stars in the
green strip and its corresponding
sequence must be different at a given luminosity.

Finally in this section, we return to the exceptional cases of (1) the
green strip originating from sequence B which lies on the
long period side of sequence C$^\prime$ and (2) the green strip
originating from sequence C$^\prime$ which lies on the short period
side of sequence B.  Remember that in Section~\ref{pr1}, it was deduced
that the $\ell=1$ mode was the dominant mode in sequence A and that
this mode has a period which is $\sim$0.043 longer in $\log P$ than
the radial mode with $\ell=0$.  Recall also that there appear to be at
least two modes (of the same radial order) associated with sequence B.
We will show later that these are most likely the $\ell=1$ and $\ell=0$
modes, with the former being most common.
Finally, as noted in Section~\ref{seqs}, stars on sequence C appear
to be pulsating in the radial fundamental mode.  With these facts
in mind, the exceptional cases noted here can be explained naturally
if stars on sequences A$^\prime$, A and B are most commonly pulsating
in the $\ell=1$ mode in both their primary pulsation mode and their
secondary modes $P2$ and $P3$, while stars on sequences C$^\prime$ and
C are most commonly pulsating in the radial $\ell=0$ mode in both
their primary pulsation mode and their secondary modes.  Thus the
stars with primary modes on sequence B will have $\log P$ for their
secondary modes increased relative to $\log P$ of the radial mode
pulsators on sequence C$^\prime$ by $\Delta\log P \sim 0.043$, thereby
causing the exceptions noted in points (1) and (2) above.  Note that
the offset $\Delta\log P \sim 0.043$ is derived from the populous RGB.
Since the slope of sequences C and C$^\prime$ is steeper than the
slope of sequences A$^\prime$, A and B, this offset will increase with
luminosity and will be larger for AGB stars.  

An alternative explanation for the exceptional cases would be for the
stellar mass to decrease from sequence A$^\prime$ to A to B, then to
increase to sequence C$^\prime$ then decrease again to C.  This seems
contrived and is not consistent with the evolutionary picture
described in the next section.

\subsection{The stars on sequence D}

Although generally not part of this study, in the last panel of 
Figure~\ref{channel} the secondary periods of stars with their
primary period on sequence D are shown as green points.  This plot highlights the
fact that the periods of pulsation associated with the LSPs
of sequence D lie mainly on or near sequence B.

\subsection{The stars on sequence F}

The faint sequence F between sequences C$^\prime$ and C is marked in
Figure~\ref{wjk-p}.  In panel C$^\prime$ of Figure~\ref{channel}, it
can be seen that secondary periods of stars whose primary period lies
on sequence C$^\prime$ form a sequence close to sequence F.  This
suggests that the sequence F stars are possibly fundamental mode
pulsators like the stars on sequence C.  However, it is not clear why
the sequence F stars should form a distinct sequence.  One possible
explanation is that they are high mass LPVs (so that their periods are
shorter than the periods typical of sequence C stars) formed in a
relatively recent burst of star formation.

\section{The evolution of LPVs}

\subsection{Tracks in the PL diagram}\label{evsec}

\begin{figure}
\includegraphics[width=1.0\columnwidth]{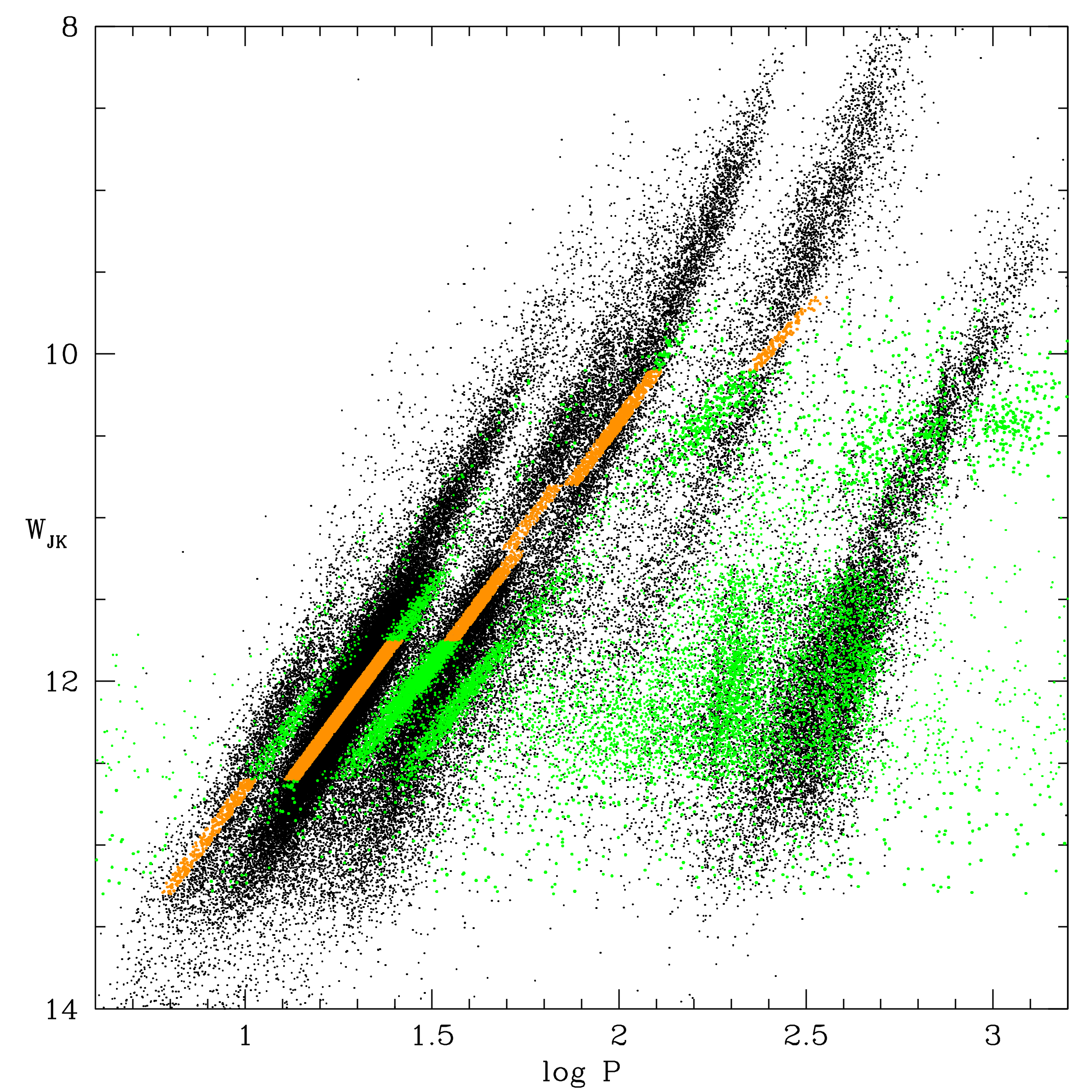}
\caption{
Plots of $W_{\rm JK}$ against the primary period $P1$ for stars in the
OGLE III catalogue (black points).  Evolutionary
channels cross each sequence from left to right and stars
whose primary periods fall in these channels are plotted as orange
points.  The secondary periods $P2$ and $P3$ of the
stars with $P1$ in the orange channels are plotted as green
points.  See text for further details.  The colours can be seen 
in the online version of the paper.
}
\label{evol}
\end{figure}

For most of the stars in PL diagrams such as Figure~\ref{wjk-p}, mass
loss is small so that the evolutionary track of a star is represented
by a line of constant mass and pulsation mode.  An examination of
pulsation models shows that in the ($W_{\rm JK},\log P$) plane, lines
of constant mass and pulsation mode have a smaller slope than the
sequences \citep[e.g.][]{sos13a}.  In Figure~\ref{evol}, channels
similar to those in Figure~\ref{channel} have been drawn but now the
channels are roughly parallel to lines of constant
mass and pulsation mode so they represent evolutionary tracks.
The slopes of the tracks are estimated from the
pulsation models described in Section~\ref{comp_mods}.  For
convenience of plotting, at lower luminosities on sequences 
A$^\prime$, A and B, the orange channels have been drawn more 
tilted relative to the sequences than the models indicate but this 
does not qualitatively affect the empirical evolutionary 
scheme presented here.

Consider stars in Figure~\ref{evol} whose primary period lies in the
orange strip through sequence A. As stars in this channel evolve to
higher luminosity, they slowly move from the short period side of
sequence A to the long period side.  At the same time, there is a
declining prominence of the secondary period corresponding to the next
highest radial order (the green strip to the left of the orange
channel in sequence A) and an increasing prominence of the secondary
period corresponding to the next lowest radial order (the green
strip to the right of the orange channel in sequence A).  This means
that the primary period most likely detected by observations will move
to a lower radial order as the luminosity increases.  Thus the orange
points representing the evolving stars in Figure~\ref{evol} will jump
from the long period side of sequence A to the short period side of
sequence B because the highest amplitude mode has moved from that
associated with sequence A to that associated with sequence B.
Evolution will then continue up and across sequence B before a jump to
sequence C$^\prime$ and so on.  Note that 
in the empirical scheme shown here, the point 
of transition of the orange channels from sequence to sequence is 
judged by eye since current pulsation models
do not give reliable theoretical estimates of mode amplitudes.

When a star reaches the RGB tip, it will transit to a
lower-luminosity core-helium burning phase before rising again up
the AGB with a slightly shorter period than it had at the same
luminosity on the RGB.  This situation is shown in Figure~\ref{evol}
for the stars in the channel on sequence B.  The AGB evolution is
shown only after the RGB tip luminosity ($W_{\rm JK} \sim 11.3$)
is exceeded.  

\subsection{Mass variations through the sequences}\label{mvarinseq}

The evolutionary scenario described in the previous sub-section has
implications for the masses of stars on the sequences.  Note firstly that, on a
given sequence, stars of different mass will follow essentially
parallel tracks but the higher mass stars will be found at higher
luminosities.  This means that the mass increases up a sequence, it
increases to shorter periods across a sequence at a given luminosity,
and it also increases from sequence to sequence at a given luminosity
as the radial order increases (the sequence period decreases).  Note
that a mass difference between sequences was invoked above in order to
explain why the period ratios between the sequences did not match the
period ratios in individual stars.  The evolutionary scenario provides
an understanding for this empirical finding.

\begin{figure}
\includegraphics[width=1.0\columnwidth]{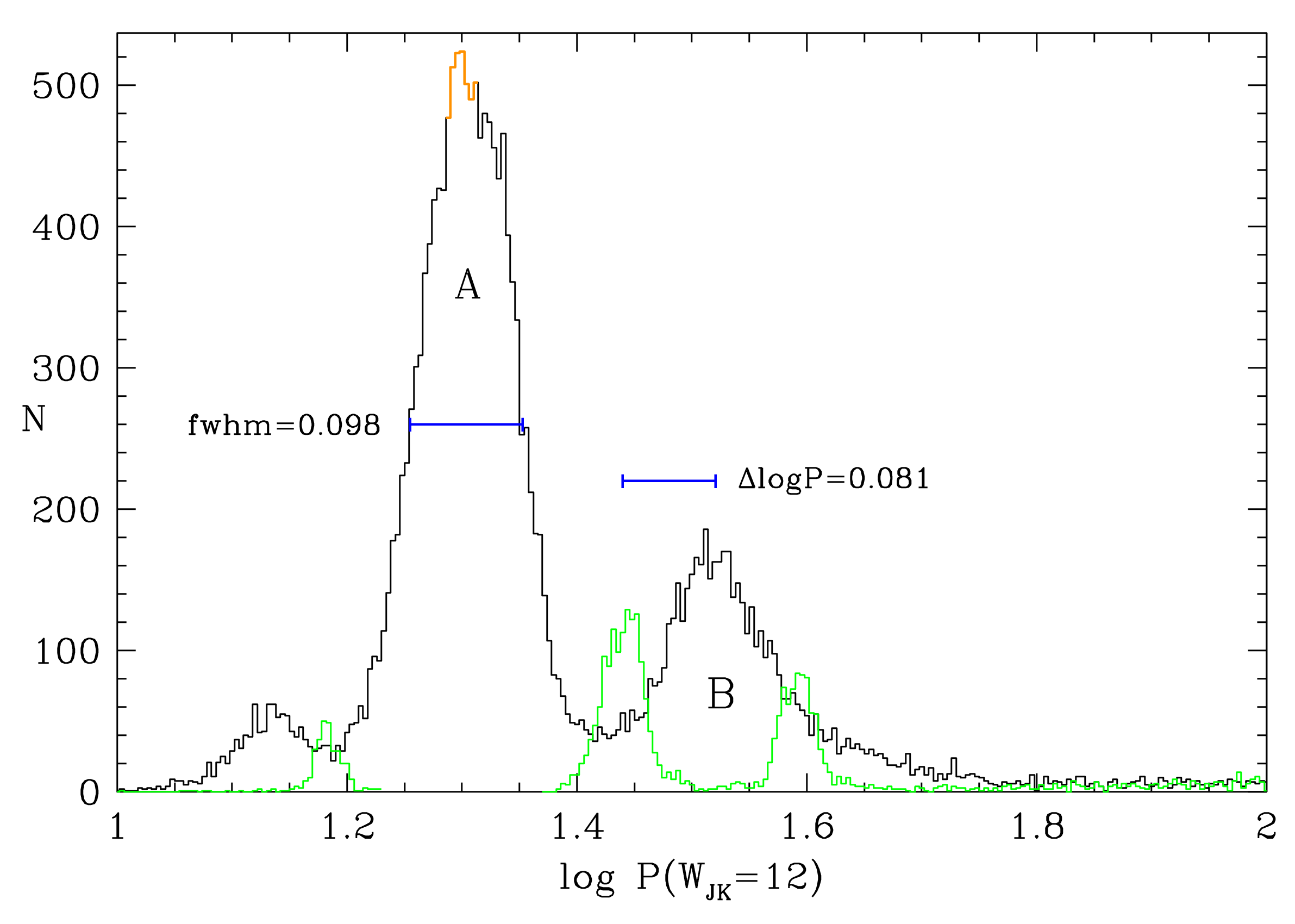}
\caption{
Histogram of the number of stars in a horizontal band defined by
$W_{\rm JK} = 12\pm0.25$ in panel A of
Figure~\ref{channel}.  The horizontal position of each star is given
by the value of its coordinate $\log P$[$W_{\rm JK}$=12].  The black curve
is obtained by using the primary period $P1$ of each star.
The orange segment of this curve represents the sequence A stars in
the orange channel through the middle of sequence A.  The
green curve is obtained by using the secondary periods $P2$ and $P3$
of those stars in the orange channel (i.e. this is the histogram of the
green points in panel A of Figure~\ref{channel}).
The colours can be seen in the online version of the paper.
}
\label{histA}
\end{figure}

The periods of stars in the various sequences formed in PL diagrams by
using the primary or secondary periods can be used to estimate the
mass variation from sequence to sequence and the mass variation across
sequences.  Figure~\ref{histA} shows a histogram of the number of
stars in a horizontal band defined by $W_{\rm JK} = 12\pm0.25$ in 
panel A of Figure~\ref{channel}.  Here we are interested in
exploring the mass variation from sequence B to sequence A and the
mass variation across sequence A.  The positions of sequences A and B
are marked on Figure~\ref{histA}.  The peak in the green
distribution at $\log P$[$W_{\rm JK}$=12] $\approx$ 1.44 is formed by
sequence A stars whose secondary periods are produced by the pulsation
mode which produces sequence B.  The difference $\Delta\log P \sim
0.081$ in the positions of this peak and the sequence B peak is caused
by the different masses of stars on sequences A and B.  Pulsation models  
for low radial order modes at the luminosities and masses of the stars
involved here show that, at a given luminosity, $P \propto M^{-0.8}$
(this result comes from the pulsation models
described in Section~\ref{comp_mods}).
Thus the shift $\Delta\log P \sim 0.081$ corresponds to the average
mass of stars on sequence A being approximately 26\% larger than
the average mass of stars on sequence B.

Mass variations \emph{within} a sequence are also required.
These mass variations make a large 
contribution to the finite width in $\log P$ of the sequences, 
with the mass increasing to shorter periods
across the sequences.  Let us consider sequence A.
The total width of this sequence (represented by
the full-width at half maximum) is $\Delta\log P \sim 0.098$.  As well
as the mass variation across the sequence, there is also a
contribution to the sequence width from the period variability
represented by the vertical width of the sub-ridges in the top panel
of Figure~\ref{mode_amps}.  This contributes approximately $\Delta\log
P \sim 0.01$ to the observed width of sequence A, leaving the
contribution from mass variation alone at $\Delta\log P \sim 0.088$.
The sub-sequences of sequence A (i.e. the multiple modes) will also
make a small contribution to the width but we ignore this.  The width
$\Delta\log P \sim 0.088$ means that the mass increases by about 29\%
from the longer period to the shorter period side of sequence A.

In the evolutionary scenario described above, the mass variation
across a sequence should be equal to the mass variation between
sequences.  Remember that stars jump between sequences from the long
period edge of one to the short period edge of the other.  Thus the
masses of stars on the long period side of a sequence should be the
same as the masses of stars on the short period edge of the adjacent
longer period sequence.  Consequently, the mass difference between
the short period edges of two adjacent sequences will be the mass
difference across the shorter period sequence.  The results given
above for sequences A and B show that the mass difference between
these two sequences is indeed close to the mass difference across
sequence A.

The procedure applied above to determine the difference in the average
masses of stars on sequences A and B can be applied to other pairs of
sequences in which stars pulsate mostly in the same mode.  We find 
that stars on sequence A$^\prime$ are $\sim$17\%
more massive than stars on sequence A at luminosities around 
$W_{\rm JK} = 12$.  The difference in the average masses of stars on
sequences C$^\prime$ and C cannot be readily determined at this
luminosity since the mode of sequence C is rarely seen there.  However,
near $W_{\rm JK} = 10.5$, which is on the AGB, many stars exhibit
primary and secondary oscillations in both of the modes associated
with sequences C$^\prime$ and C.  The period shifts indicate that the
stars on sequence C$^\prime$ are $\sim$16\% more massive than those on
sequence C.  To derive this result we used the formula
$P \propto M^{-1.1}$ rather than $P \propto M^{-0.8}$ which is appropriate at lower
luminosities on the RGB (these formulae are derived from the pulsation
models described in Section~\ref{comp_mods}).   Finally, we consider sequences B and
C$^\prime$.  Comparing the masses of stars on these two sequences 
is difficult at any luminosity because of the paucity of
stars that exhibit both the sequence B and C$^\prime$ modes, and
because the stars with primary periods on sequence B appear to be
$\ell=1$ pulsators while stars with primary periods on sequence
C$^\prime$ appear to be $\ell=0$ pulsators.  We do not make any
estimate of the mass difference between these two sequences.

\section{The luminosity dependence of period ratios}\label{ldep_pr}

\begin{figure*}
\includegraphics[width=0.92\textwidth]{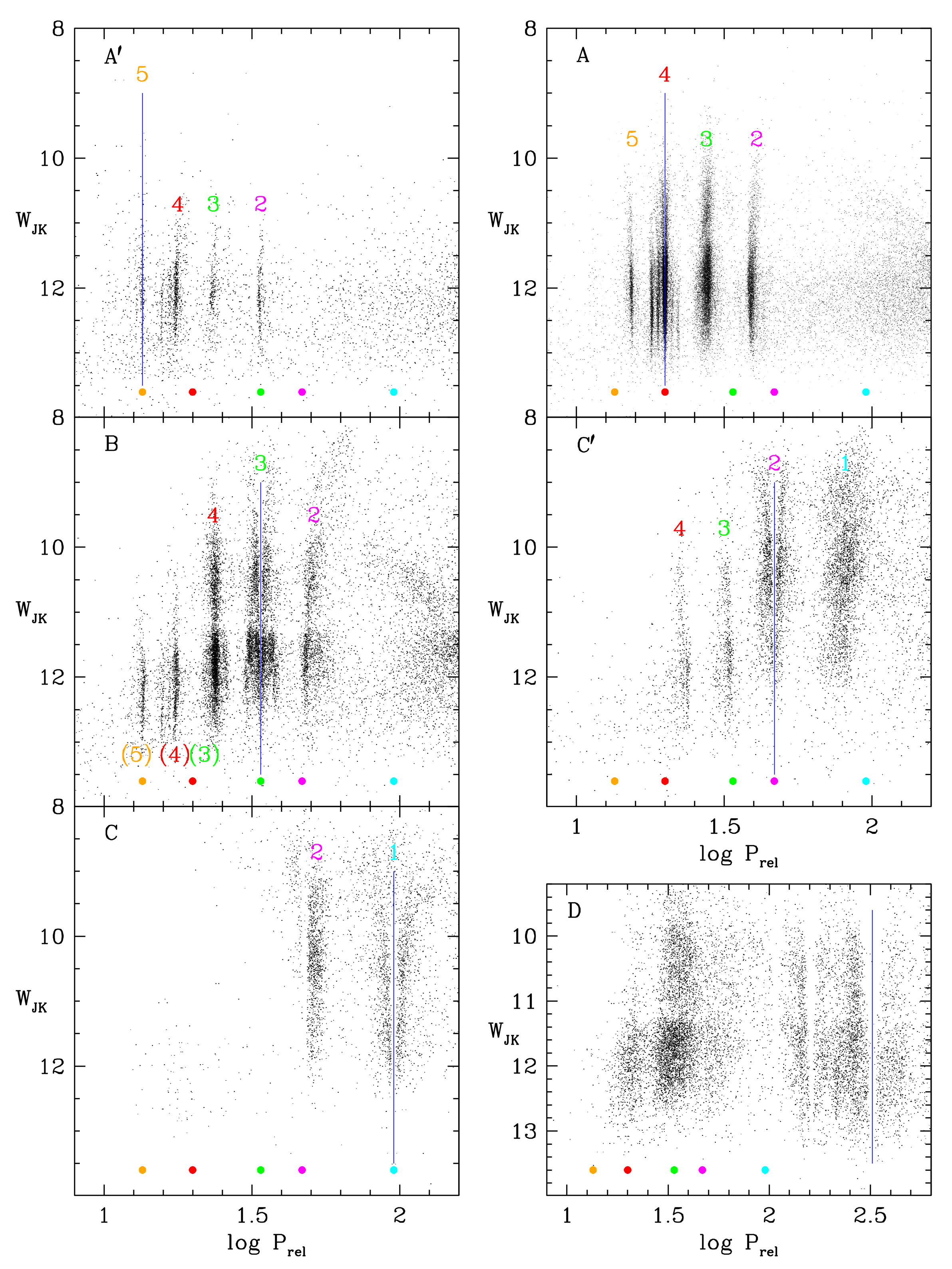}
\caption{
$W_{\rm JK}$ plotted against the ratio of secondary period to the
primary period for stars whose primary periods lie in a strip of width
$\Delta \log P1 = 0.12$ down the centre of each sequence.  Here $\log
P_{\rm rel}$ is defined as $\log P_{\rm rel} = \log (P2/P1) + \log
P_{\rm centre}$ or $\log P_{\rm rel} = \log (P3/P1) + \log P_{\rm
  centre}$, where $\log P_{\rm centre}$ is $\log P$ at the centre of
the sequence when $W_{\rm JK} = 12$.  The blue vertical line in the
panel for each sequence is situated at $\log P_{\rm centre}$.  The
coloured dots indicate $\log P_{\rm centre}$ for each sequence using
the colour coding for the sequences adopted in Figure~\ref{em_strips}.
The coloured numbers count the observed radial orders in each panel
and these numbers are aligned consistently
from panel to panel, starting at 1 for the mode with the longest 
observed period in panel C.  
The colours can be seen in the online version of the paper.
}
\label{lum_dep}
\end{figure*}

Although Figure~\ref{em_strips} demonstrates the existence of
secondary periods in many LPVs, it does not indicate how the presence
of secondary periods varies with luminosity.  The luminosity
dependence of the secondary periods can be seen in
Figure~\ref{lum_dep}.  For each sequence, this figure plots $W_{\rm
  JK}$ against the secondary period relative to the primary period
whose position is marked by a vertical blue line.  The vertical strips
of black points show where individual secondary modes are present in the
stars.  The radial order of the modes associated with these secondary
modes are indicated by coloured numbers, with radial order 1
corresponding to the fundamental mode (see Section~\ref{comp_mods} for
the justification of the assignment of radial orders).  The positions
of the individual sequences are indicated by coloured dots in each
panel.  The fact that the dots and numbers of the same colour do not
align horizontally (except for the reference mode) is yet another
demonstration of the fact that the ratio of the periods in individual
stars do not match the ratio of the periods of the sequences.

Panel A$^\prime$ of Figure~\ref{lum_dep} shows the
ratio of secondary to primary period for stars whose primary period
lies on sequence A$^\prime$.  Looking at this panel, we see three
closely-spaced vertical bands under and slightly left of the red 4.  
These bands are caused by the three modes occurring on sequence A.  The longest
period mode (the $\ell=1$ mode) becomes dominant as the luminosity
increases.  This result is also seen for stars with their primary
period on sequence A (see panel A).  There are five vertical
bands associated with sequence A and these are caused by the the three modes
present within this sequence.  As discussed in Section~\ref{pr1}, 
the shortest period band is produced
by stars whose primary mode has $\ell=1$ and whose secondary mode has
$\ell=0$ while the longest period band is produced by stars whose
primary mode has $\ell=0$ and whose secondary mode has $\ell=1$.
The shortest period band extends to higher luminosities than the
longest period band which suggests that the $\ell=1$ mode is more
prominent as the primary period (i.e. the one with largest amplitude)
at higher luminosities.  Once again, we note that the dominance of the
$\ell=1$ mode at higher luminosities on sequence A is consistent with the findings
of \citet{ste14} from Kepler observations but different from the
findings of \citet{mos13} who suggest that the radial $\ell=0$ mode
becomes dominant at high luminosities.

The stars with their primary mode on sequence B show
three\footnote{The apparent splitting near $W_{\rm JK} = 12$ of the
outer bands of sequence B are cause by the 1-year alias of the
primary period, so this feature is not real.  A similar 1-year alias
can be seen in the bands of sequence A near $W_{\rm JK} = 11.2$ if
the figure is expanded.}  vertical bands associated with the primary
mode (panel B of Figure~\ref{lum_dep}).  This structure indicates the
presence of two close modes but they seem to be confined to RGB stars
only ($W_{\rm JK} > 11.3$).  Only one mode appears to be present
on the AGB. 

Panel B in Figure~\ref{lum_dep} reveals the presence of
apparent extra modes not present in any other panel.  In a simple
sequential assignment of radial orders, there appear to be secondary
modes present corresponding to radial orders 5 and 6.  However, looking at
panel A$^\prime$ immediately above panel B, it can
be seen that the mode structure associated with the radial orders 4 and 5 in panel A$^\prime$ 
is reproduced exactly in panel B.  Another point to notice
is that the horizontal position of radial order 2 in panel A$^\prime$ 
coincides exactly with the position of sequence B.  We conclude from
these facts that stars whose primary mode lies on sequence B are of
two types: (1) stars pulsating in the third radial order with secondary
periods in the second and fourth radial orders, and (2) stars
pulsating in the second radial order with secondary periods in the third, 
fourth and fifth radial orders.  The type 2 stars are more massive than
type 1 stars and they have masses typical of stars on sequence A$^\prime$.  These
type 2 stars are really stars belonging to sequence A$^\prime$ but the
observed amplitude of the radial order 2 pulsations is larger than
the observed amplitude of the radial order 5 pulsations so they end
up on sequence B.  The apparent extra radial orders in panel B have
been labelled by their true radial orders using numbers in brackets.

If we consider the stars on sequence B whose primary mode is of radial
order 3 (panel B of Figure~\ref{lum_dep}), we see that 
the secondary modes pulsating in radial order 4
form two bands with the shorter
period band being overwhelmingly dominant (the apparent band on the
short period side of the radial order 4 group belongs to radial order 3
of the type 2 stars examined in the last paragraph).  Given that there
are two possible primary modes of radial order 3, the dominance of
the shorter period band in panel B suggests that the dominant primary
mode is the longer period one.  The separation $\Delta \log P \approx 0.04$ 
of the two modes associated with radial order 3 (derived from the separation
of the vertical bands associated with this order in panel B of Figure~\ref{lum_dep})
is consistent with the separation of the $\ell =0$ and
the $\ell =1$ modes of radial order 3 in the models of \citet{ste14}.
These results therefore suggest that the $\ell =0$ and
the $\ell =1$ modes are present on sequence B and that the $\ell=1$ mode is dominant.
The direct finding that the $\ell=1$ mode is dominant on both sequences B and A is
consistent with the suggestion in Section~\ref{chans} that the $\ell=1$ mode
is dominant on each of sequences B, A and A$^\prime$.

\section{Comparison of observations and radial pulsation models}\label{comp_mods}

\begin{figure*}
\includegraphics[width=0.93\textwidth]{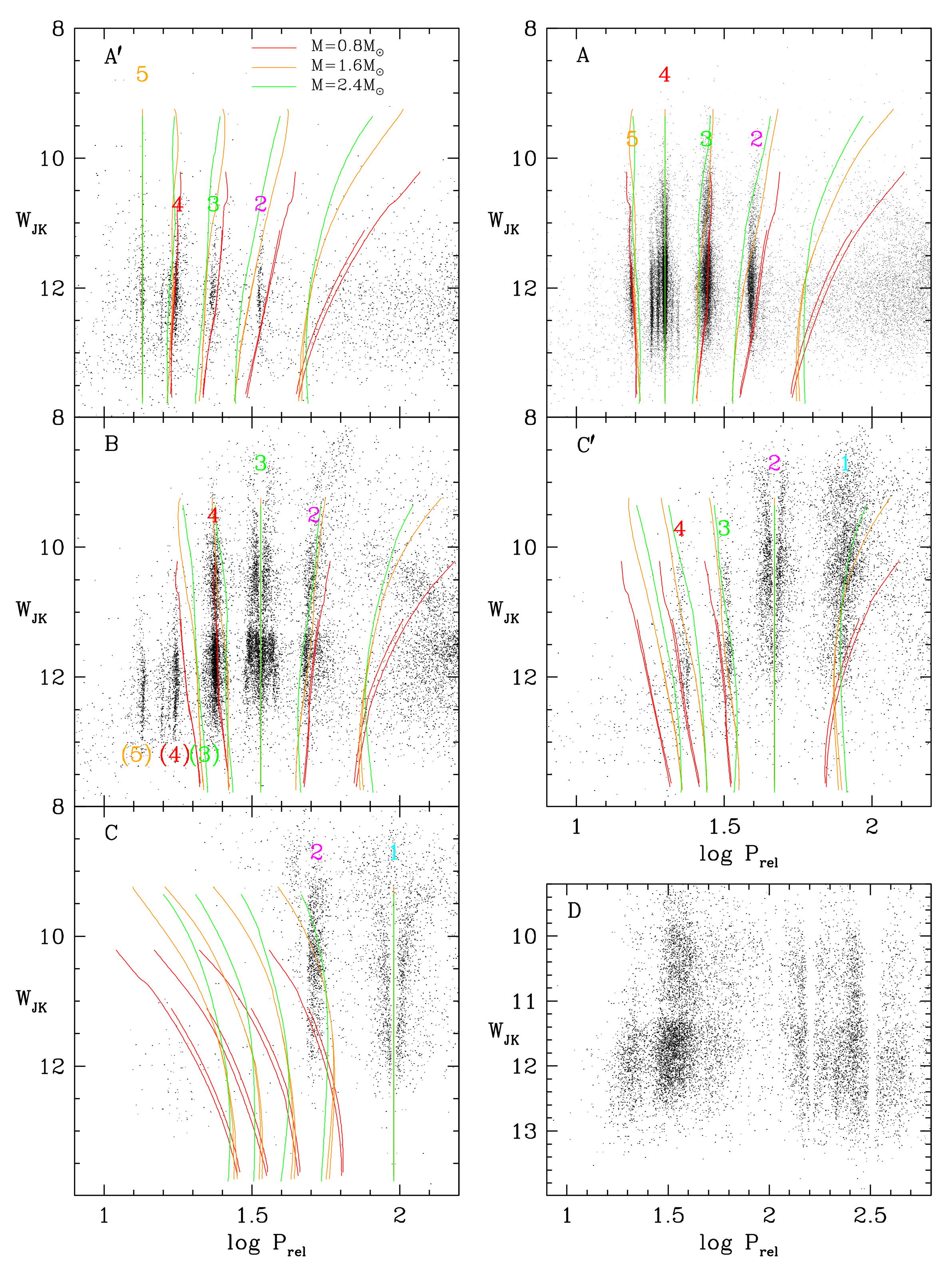}
\caption{
Same as Figure~\ref{lum_dep} with the addition of radial pulsation models showing the
logarithmic period ratios for the first five modes.  It is assumed that 
sequence C corresponds to the fundamental mode (radial order 1)
and that sequences C$^\prime$ to A$^\prime$ correspond to the first to fourth overtone
(radial orders 2 to 5), respectively.  In each panel, the model period ratios
are taken relative to the radial order appropriate for the sequence assigned to the panel.
Models with masses of 0.8, 1.6 and 2.4\,M$_{\odot}$
are represented by red, orange and green lines, respectively.
Both AGB and RGB models are plotted for $M$ equal to
0.8 and 1.6 M$_{\odot}$ (there is no RGB for $M = 2.4$\,M$_{\odot}$).
The AGB and RGB models are only separable for the fundamental mode.
The colours can be seen in the online version of the paper.
}
\label{lum_dep+mods}
\end{figure*}

\begin{figure*}
\includegraphics[width=0.95\textwidth]{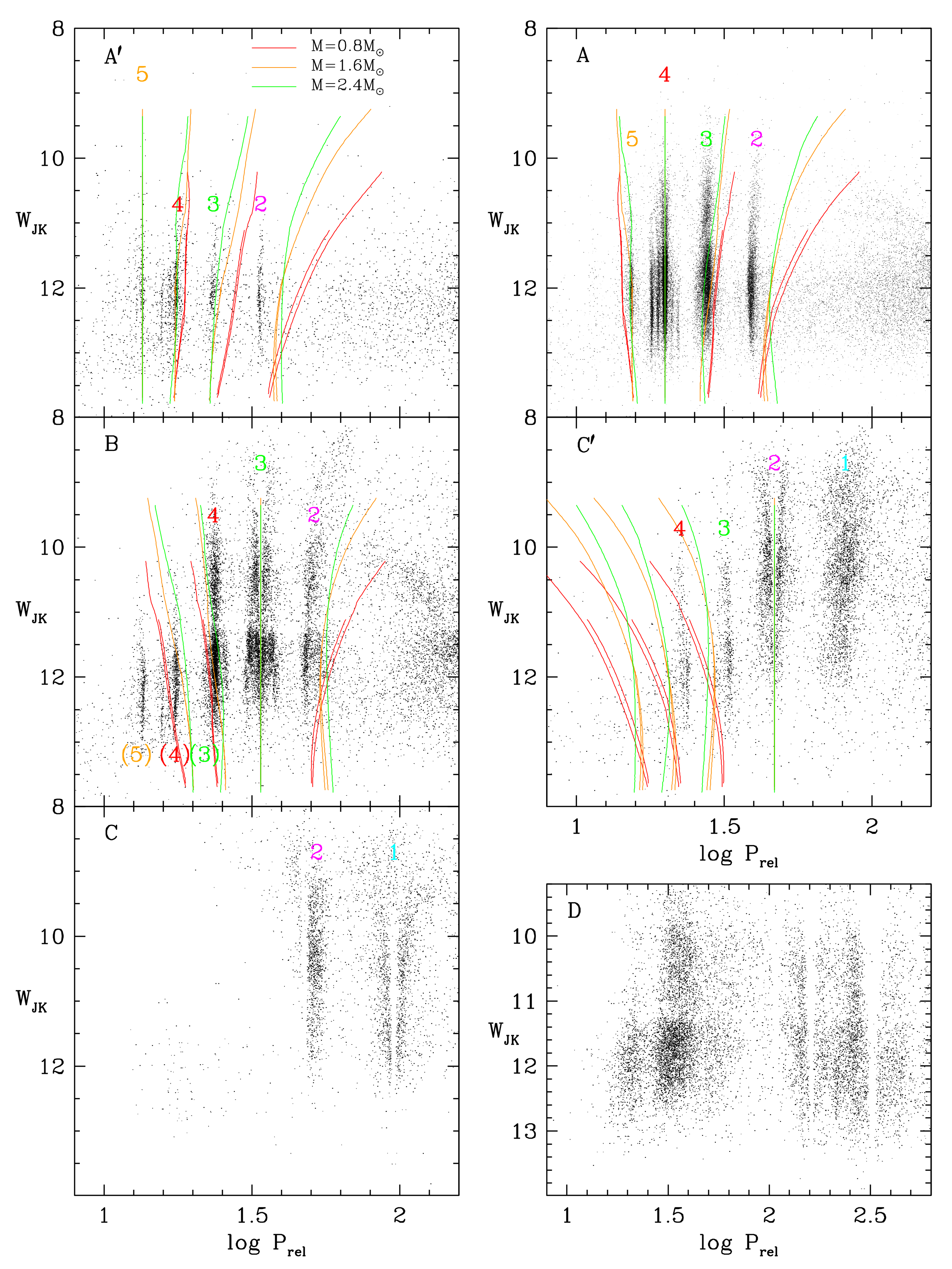}
\caption{
Same as Figure~\ref{lum_dep+mods} but now it is assumed that
sequence C$^\prime$ corresponds to the fundamental mode (radial order 1)
and that sequences B, A and A$^\prime$ correspond to the first to third overtone
(radial orders 2 to 4), respectively.  Only the first four radial modes of the
models are shown.
}
\label{lum_dep+wrong_mods}
\end{figure*}

In this section, period ratios observed in individual stars are compared
to period ratios in radial pulsation models of red giants.  
The linear, non-adiabatic pulsation models
were computed with the code described in \citet{wo14} using the 
same input parameters (abundances, mixing-length, turbulent viscosity 
parameter, core mass) as in that paper.
Evidence was presented in Section~\ref{chans} that, in a given star,
both the dominant primary pulsations and the secondary pulsations
correspond to modes of the same spherical degree $\ell$.  Thus,
although the nonradial $\ell=1$ mode appears to dominate the primary
pulsations in sequences A$^\prime$, A and B, provided the ratio of the
$\ell=1$ to the $\ell=0$ periods is approximately the same for radial
orders 3 to 5 associated with these sequences, the observed period
ratios in individual stars should be the same as the period ratios in
radial pulsation models.  In Figures~\ref{lum_dep+mods} and
\ref{lum_dep+wrong_mods} the observed period ratios in individual
stars are compared to the period ratios in radial pulsation models.

In Figure~\ref{lum_dep+mods}, it is assumed that the radial mode
associated with sequence C (the Mira sequence) is the fundamental mode
(radial order 1) and that sequences C$^\prime$, B, A and A$^\prime$
correspond to successively higher radial orders of pulsation.  It can be seen
that for each of the panels in Figure~\ref{lum_dep+mods}, the model
period ratios match the observed period ratios well for some mass
in the range 0.8 to 2.4\,M$_{\odot}$ except for the highest mass
and luminosity models in panels C$^\prime$ and C. This problem is
discussed in Section~\ref{model-mass}.  Note that in panel B, there is a
mismatch between the models and those vertical bands that really belong to
sequence A$^\prime$ (as described in Section~\ref{ldep_pr})
but this is expected because of the different masses
of stars on sequences A$^\prime$ and B.  Overall,
these results provide strong evidence that the primary pulsation
modes associated with sequences C, C$^\prime$, B, A and A$^\prime$
belong to radial orders 1, 2, 3, 4 and 5.  It also provides strong
evidence that both primary and secondary pulsations in a given star
tend to belong to the same $\ell$ value.

In Figure~\ref{lum_dep+wrong_mods}, it is assumed that the radial mode
associated with sequence C$^\prime$ is the fundamental mode (radial
order 1) and that sequences B, A and A$^\prime$ correspond to
successively higher radial orders.  Clearly, the mode marked by the
cyan number 1 (sequence C) has no explanation in terms of radial modes with
this assignment of radial orders.  Furthermore, even for higher radial
orders, the ratios of mode periods in the models fail to match the
observed ratios.  We therefore reject this modal assignment.

\subsection{Comparison with previous modal assignments}

The finding here that sequences C, C$^\prime$, B, A and A$^\prime$
correspond to radial orders 1, 2, 3, 4 and 5 is consistent with most
earlier findings \citep[e.g.][]{woo96,woo99,tak13}.  However, this
assignment is inconsistent with the suggestion of 
\citet{sos07} and \citet{mos13} that sequence C$^\prime$
corresponds to radial order 1.

In terms of nonradial modes, we find direct evidence for the presence
of the $\ell =1$ mode on sequences A and B and for the $\ell=2$
mode on sequence A.  In their analysis of OGLE III data, \citet{tak13}
claim to find the presence of the two nonradial modes. Firstly, they
associate the $\ell =2$ mode of radial order
three\footnote{\citet{tak13} use a nomenclature in which the radial
order is offset by one from the nomenclature used here.  For
example, their $p$-mode $p_2$ corresponds to radial order 3 in this
paper.} with sequence B while the only nonradial mode we find has $\ell =1$.
Secondly, they find the $\ell =1$ mode of radial order 5 associated
with sequence A whereas we associate radial order 5 with sequence
A$^\prime$ and we find the $\ell =1$ mode of radial order 4 associated
with sequence A.  Because \citet{tak13} only looked at LPVs on the RGB
designated as OSARGS in the OGLE III catalogue, all LPVs with periods on
sequences A$^\prime$ and C were excluded (by definition, OSARGS on the
RGB do not have these modes) so they had an incomplete picture of the
modes involved.  In addition, the Petersen diagram they used meant
that a number of modes overlapped on the figure causing confusion.

\subsection{The mass dependence of the period ratios}\label{prat-m}

In the lower panel of Figure~\ref{em_strips}, the various clusters of
points representing period ratios of specific pairs of modes tend to
form ridges that are not horizontal.  For example, we now know that
the ridge at ($\log P1$[$W_{\rm JK}$=12], $\log Pi$/$P1$) $\approx$
(1.3, 0.3) corresponds to stars whose primary pulsation mode belongs
to radial order 4 and whose secondary pulsation mode belongs to radial
order 2.  This group of points, and most of the other groups of
points, form ridges such that the period ratio becomes smaller when
moving from the long period side of the primary mode sequence to the
short period side.  It was shown in Section~\ref{mvarinseq} that the
mass of stars in a sequence increases when moving from the long period
side to the short period side of the sequence.  This suggests that the cause of the
non-zero slope of the ridges in the lower panel of Figure~\ref{em_strips} is
mass variation across sequences.  The models shown in
Figure~\ref{lum_dep+mods} confirm this.  If we take the example given
above for the ridge at ($\log P1$[$W_{\rm JK}$=12], $\log Pi$/$P1$)
$\approx$ (1.3, 0.3), panel A of Figure~\ref{lum_dep+mods} shows that
the period ratio of radial order 2 to radial order 4 modes decreases
as the model mass increases, just as suggested by the slope of the
ridges.

It was noted in Section~\ref{pr1} that the ratio of the periods of the
$\ell=1$ and $\ell=0$ modes of radial order 4 increased slightly
from the long period side to the short period side of sequence A (the
opposite slope to the cases considered in the previous paragraph).  This
result provides evidence for the mass dependence of the period ratio
of the $\ell=1$ and $\ell=0$ modes, with the period ratio increasing
with mass.  Combining the results for the variation
of the period ratio across sequence A from Section~\ref{pr1} with
the results for the variation of mass across sequence A from 
Section~\ref{mvarinseq}, we find that 
$\partial\log (P(\ell=1)/P(\ell=0))/\partial\log M \approx 0.036$ 
for the stars on sequence A.  This result provides a test for
nonradial pulsation models of luminous red giant stars.

\subsection{Pulsation masses from the models}\label{model-mass}

\begin{figure}
\includegraphics[width=1.0\columnwidth]{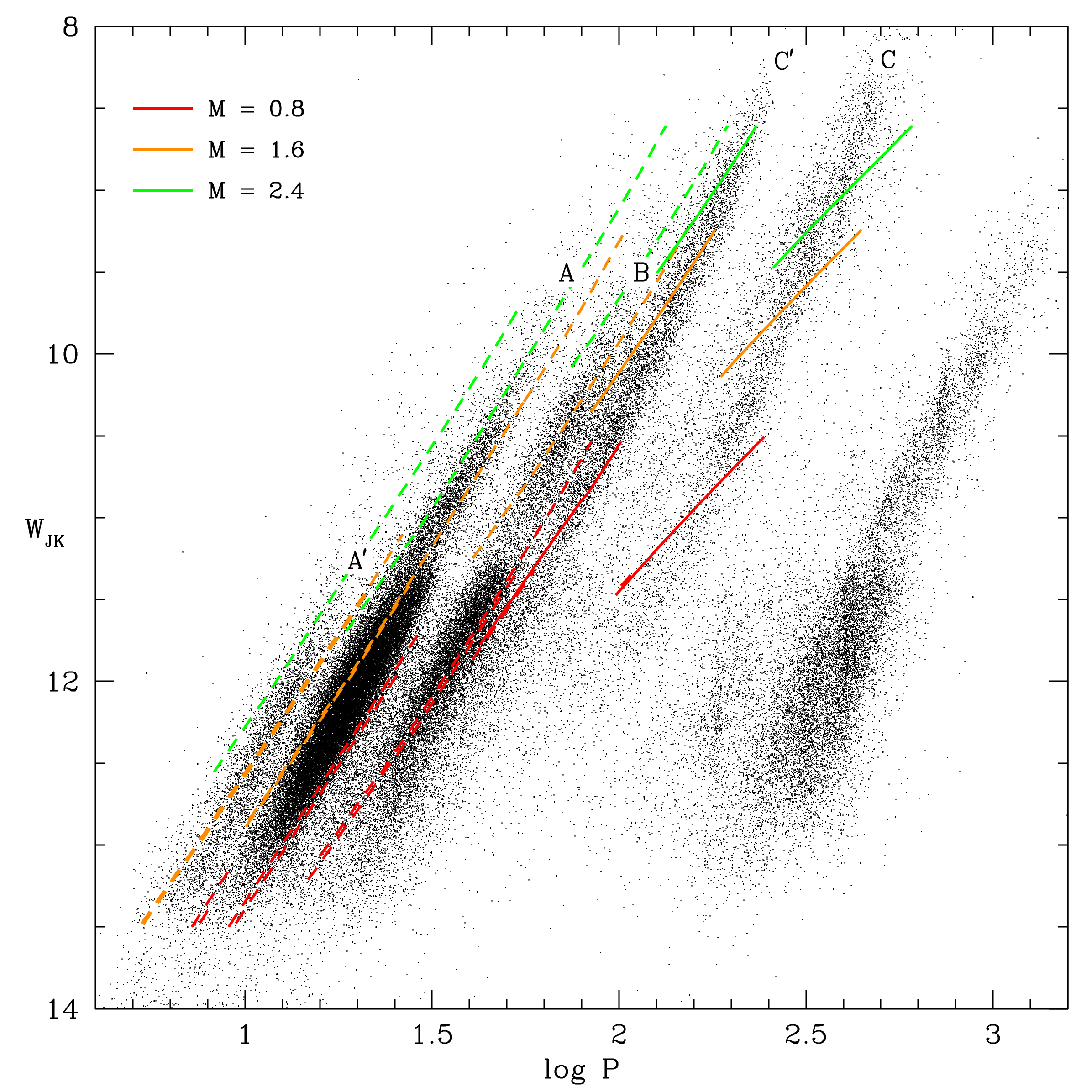}
\caption{
Plots of $W_{\rm JK}$ against the primary period $P1$ for stars in the
OGLE III catalogue (black points). The coloured lines show the positions
of modes from the first five radial orders for stars of mass 0.8, 1.6
and 2.4 M$_{\odot}$.  The lines for each mode are plotted only in regions where they
cross the sequence in which the mode is expected to dominate.  The first two
radial orders are shown as continuous lines and they correspond to
the fundamental and first overtone radial pulsation modes.  The lines
for the next three radial orders are shown as dotted lines because
the periods shown have been increased by $\Delta\log P = 0.043$
from the period of the radial modes in an attempt to simulate the
periods of the $\ell = 1$ pulsation mode.
The colours can be seen in the online version of the paper.
}
\label{masses}
\end{figure}

Until now in this paper, only period ratios have been compared
to models because period ratios are less sensitive to model parameters
than are the periods themselves.  In addition, to compare
model and observed periods in (say) the ($W_{\rm JK}$, $\log P$) plane,
conversions from model $T_{\rm eff}$ and $L$ to the observed
$W_{\rm JK}$ are required.  These conversions are not particularly well
known for cool stars such as LPVs and there are additional problem
due to interstellar and circumstellar reddening and the distance modulus
(although the latter is now well-determined for the LMC).  Keeping
these problems in mind, in this
final sub-section, we do compare the models to the observations in the
($W_{\rm JK}$, $\log P$) plane.

For each stellar model, $T_{\rm eff}$ and $L$ have been converted to
$J$ and $K$ in the 2MASS system using the tabulated bolometric corrections
obtained from the model atmosphere calculation by \citet{hou00a} and
\citet{hou00b} for a metal abundance $\rm{[Fe/H]} = -0.5$ which is appropriate
for the LMC and which is consistent with the metal abundance used in the model
calculations.  Conversion of $J$ and $K$ in the system of
\citet{hou00a} and \citet{hou00b} to the 2MASS system is done using the
conversions in \citet{car01}.  An LMC distance modulus of 18.54
and a reddening $E(B-V)$ of 0.08 from \citet{kw06} are used, along
with the reddening law of \citet{car89} to get the reddening at $J$ and $K$.

Figure~\ref{masses} is a plot in the ($W_{\rm JK}$, $\log P$) plane
of the periods and luminosities of modes from the
first five radial orders of stars with masses of 0.8, 1.6 and 2.4 M$_{\odot}$.
For each mode, only a short line segment is plotted centred on the position where it crosses
the sequence in which the mode is assumed to be dominant.  Note that the models
computed here only produce periods for radial modes.  However, it was shown 
in Sections~\ref{chans} and \ref{ldep_pr} that the $\ell = 1$ mode is
dominant on sequences A$^\prime$, A and B, and in Section~\ref{comp_mods}
it was shown that these sequences correspond to radial orders 5, 4 and 3, respectively.
Furthermore, in Sections~\ref{pr1} and \ref{chans} it was shown that the
period of the $\ell =1$ mode is longer than the period of the $\ell =0$ mode
by $\Delta\log P \approx 0.043$.  Hence, to compare the models with the
observations, the computed periods for radial orders 3, 4 and 5 
have been increased by $\Delta\log P = 0.043$ before being plotted 
in Figure~\ref{masses}.  This
correction is made in order to
simulate the $\ell=1$ mode periods.  The resulting periods 
are shown in Figure~\ref{masses} by dashed lines.

We now make a detailed comparison of the models with the observations.  
Looking at sequence C
in Figure~\ref{masses}, it can be seen that
fundamental mode pulsation in stars with masses in the range 0.8--2.4
M$_{\odot}$ explains most of these stars, with some stars outside this
mass range required to explain the full length of the sequence (note
that the present-day population of red giant stars in the LMC peaks at
masses around 1.6 M$_{\odot}$ e.g. \citealt{nie12}).  Stars in a
similar mass range, but pulsating in the radial first overtone mode
can explain the stars on sequence C$^\prime$.  Similarly, this mass
range fits sequences $B$, $A$ and $A^\prime$ but at lower
luminosities, as expected from the empirical evolution scheme
described in Section~\ref{evsec}. The mass required to fit these three
sequences at a given luminosity increases when moving from sequence $B$ to $A$ to $A^\prime$,
consistent with the results derived in Section~\ref{mvarinseq} from the empirical evolution
scheme.  In addition, the width of sequence A requires a significant
range of masses to be present at a given luminosity.  This is also consistent with
the results derived in Section~\ref{mvarinseq}.  Given the various
results presented in this paragraph, there appears to be reasonable
agreement between the models and the observations in the ($W_{\rm
  JK}$, $\log P$) plane in spite of the potential difficulties
mentioned above.

There is one parameter on which the models and the empirical evolution scheme
are in relatively poor agreement.  This is the transition luminosity at which
stars move from one sequence to the next, especially in the case of the transition
from sequence C$^\prime$ to sequence C.  Looking at the evolutionary track
for a 1.6 M$_{\odot}$ star (i.e. the orange lines in Figure~\ref{masses}),
we see that the first overtone mode does not cross the long period edge of
sequence C$^\prime$ until $W_{\rm JK} \sim 9$.  According to the empirical evolution scheme,
the star should then jump to the short period edge of sequence C
and start pulsating in the fundamental mode at $W_{\rm JK} \sim 9$.  
However, the period of the fundamental mode of the 1.6 M$_{\odot}$ star
at this stage puts the star at the \emph{long} period edge of sequence C or beyond.
This suggests that the ratio of the fundamental to first overtone periods in the
models is too large.  Actually, the disagreement between the models
and the observations is not as bad as just suggested.  
If we look at panel C$^\prime$ of Figure~\ref{lum_dep+mods}, we see that
the logarithmic period difference $\Delta\log P$ between the 
fundamental mode to the first overtone
mode is $\sim$0.26 around $W_{\rm JK} \sim 9-10$.  This is significantly
larger than the difference $\Delta\log P \sim 0.18$ between the long period edge of
sequence C$^\prime$ and the short period edge of sequence C.  
At $W_{\rm JK} = 10$,
$\Delta\log P$ for the 1.6 M$_{\odot}$ star is indeed $\sim 0.26$ but by
$W_{\rm JK} = 9$ $\Delta\log P$ is much too large at $\sim 0.41$
so that there is still a discrepancy between the model period ratios
for the first two modes and the observed ratio of these two modes.
It is worth noting that the discrepancy discussed here seems to
be worst at high masses and luminosities.  At $M = 0.8$\,M$_{\odot}$, the 
models predict a transition from the long period edge of
sequence C$^\prime$ to the short period edge of sequence C which is
as predicted by the empirical evolution scheme.

Finally, we take a look at the period difference between RGB and AGB stars.
The lines drawn in Figure~\ref{masses} belong to both RGB
and AGB stars.  The computed separation of the RGB and AGB periods for each mode
at a given luminosity is almost imperceptible on the plot (the
separation can just be seen for the 0.8 M$_{\odot}$ models).
It is certainly much less than the period shift of $\Delta\log P \sim 0.05$ evident on sequence A
when moving from the RGB at $W_{\rm JK} > 11.3$ to the AGB at $W_{\rm JK} < 11.3$.
The observed period shift can not be explained by mass loss
between the RGB tip and the AGB since mass loss increases
the period of pulsation whereas the period is observed to decrease.  
The models thus predict that there is a very small period change
between the RGB to the AGB for a given mode.  A possible 
interpretation of the much larger period shift observed on
sequence A when going from the RGB to the AGB is that the AGB is dominated
by radial ($\ell=0$) pulsators while the RGB is dominated by nonradial
$\ell=1$ pulsators (remember that $\Delta\log P \sim 0.043$ for the period
ratio of the $\ell=1$ and $\ell=0$ modes).  However, this result conflicts with the
results in Section~\ref{ldep_pr} which suggested that the $\ell=1$ mode
becomes more dominant at higher luminosity.  The observational resolution of this
conflict awaits larger data sets such as those that will come from
OGLE IV.

\section{Summary and Conclusions}

Multiple pulsation modes in stars in the OGLE III catalogue of LPVs in
the LMC have been examined.  It has been shown that the ratios of the
multiple periods in individual stars do not agree with the ratios of
the periods of the sequences of LPVs in the PL diagram.  This leads to
the conclusion that the average masses of stars on the different
sequences vary, with the shorter period sequences containing more
massive stars.  An examation of the evolution of stars in the PL
diagram shows that the mass variation between sequences is the same as
the mass variation across individual sequences.  As an example, it is
shown that the average mass of a star on sequence A is $\sim$26\%
higher than the average mass of a star on sequence B and that the mass
variation across sequence A is of a similar magnitude, with the more
massive stars on the shorter period side of the sequence.  In general,
the masses of stars at a fixed luminosity on adjacent sequences are
found to increase by $\sim$16--26\% towards the shorter period
sequence.  Fitting pulsation models to the period ratios in multimode
pulsators shows that the sequences C, C$^\prime$, B, A and A$^\prime$
correspond to pulsation in radial orders 1 to 5, respectively (with
radial order 1 corresponding to the fundamental mode).  Closely-spaced
modes revealed by the period ratios in individual stars show the
presence of nonradial modes.  The observed period ratios indicate the
presence of the $\ell=0$, 1 and 2 modes on sequence A (radial order 4)
and the $\ell=0$ and 1 modes on sequence B (radial order 3).  The
relative populations and amplitudes of the modes of different $\ell$
show that the $\ell=1$ mode is the most common and of the largest
amplitude in both sequences.  The positions of secondary mode
sequences relative to primary mode sequences in the PL diagram also suggest
that the $\ell=1$ mode is the most prominent primary and secondary
mode on sequences A$^\prime$, A and B and that the $\ell=0$ mode is
the most prominent primary and secondary mode on sequences C$^\prime$
and C. This result is supported by the good fit of the period ratios
in radial pulsation models to the observed period ratios.

\section*{Acknowledgments}
I thank the anonymous referee whose constructive and detailed comments
have considerably improved the readability of the paper.

\label{lastpage}

\end{document}